\documentclass{pasj00}
\usepackage{natbib}

\usepackage[pdftex]{graphicx} 

\usepackage{times}

\begin{document}

\title{Thermal Tomography of the Inner Regions of Protoplanetary Disks with the ngVLA and ALMA}
\author{Satoshi Okuzumi,\altaffilmark{1} Munetake Momose,\altaffilmark{2} and 
Akimasa Kataoka\altaffilmark{3}
}
\altaffiltext{1}{%
   Tokyo Institute of Technology, 2--12--1 Ookayama,\\
   Meguro, Tokyo 152--8551}
\altaffiltext{2}{%
   Ibaraki University, 2--1--1 Bunkyo, \\
   Mito, Ibaraki 310-8512}
\altaffiltext{3}{%
   National Astronomical Observatory of Japan, 2--21--1 Osawa,\\
   Mitaka, Tokyo 181--8588}
\email{okuzumi@eps.sci.titech.ac.jp}
\KeyWords{dust, extinction --- planets and satellites: formation --- protoplanetary disks --- submillimeter: planetary systems --- magnetohydrodynamics (MHD) --- telescopes}

\maketitle

\begin{abstract}
Understanding the temperature structure of protoplanetary disks is crucial for answering the fundamental question of when and where in the disks rocky planets like our own form. However, the thermal structure of the inner few au of the disks is poorly understood not only because of lack of observational constraints but also because of the uncertainty of accretion heating processes. Here, we propose thermal tomography of the inner regions of protoplanetary disks with the ngVLA and ALMA. The proposed approach is based on the assumption that the inner disk regions are optically thick at submillimeter wavelengths but are marginally optically thin at longer millimeter wavelengths. By combining high-resolution millimeter continuum images from the ngVLA with submillimeter images at comparable resolutions from ALMA, we will be able to reconstruct the radial and vertical structure of the inner few au disk regions. We demonstrate that the thermal tomography we propose can be used to constrain the efficiency of midplane accretion heating, a process that controls the timing of snow-line migration to the rocky planet-forming region, in the few au regions of protoplanetary disks at a distance of 140 pc. 
\end{abstract}

\section{Introduction}
The temperature structure of protoplanetary disks dictates where planets of different compositions form in the disks.
Rocky planets including the Earth are generally believed to have formed inside of the snow line, where the disk temperature reaches the sublimation temperature of water ice.  
Enhancement of solids outside the snow line due to ice condensation may help gas giants form within the lifetime of the gas disks \citep{Kokubo02}. 
The composition of solids may even determine how planets form because the stickiness of solid particles generally depends on their surface composition \citep{Chokshi93}.  

Despite its importance in planet formation, the temperature structure of the inner regions of protoplanetary disks is poorly understood.  
Conventionally, the disk temperature structure has been studied based on the viscous accretion disk model that assumes vertically uniform viscosity. According to this classical disk model, accretion heating at the midplane pushes the midplane snow line in $\sim 1~\rm Myr$-old disks out to 2--3 au \citep{Davis05,Garaud07,Oka11,Bitsch15}. However, the validity of the classical viscous model has been questioned by recent magnetohydrodynamical (MHD) simulations of protoplanetary disks showing that accretion heating can only take place near the disk surface, where the ionization fraction is relatively high \citep{Hirose11,Mori19}. A recent disk temperature model based on MHD simulations predicts that the snow line in $\sim 1~\rm Myr$-old disks must lie interior to 1 au \citep{Mori21}.
This raises the fundamental question of why the Earth is so dry with a bulk water content likely well below 1 wt\% \citep{Abe00,Marty12}, because the solids outside the snow line in the solar nebula would have contained water ice in excess of 1 wt\% as witnessed by carbonaceous chondrites and comets \citep[e.g.,][]{vanDishoeck14}.
In fact, simulations show that rocky planetary embryos at 1 au are likely to grow into waterworlds if the snow line migrates inward of the 1 au orbit within 1 Myr after disk formation \citep{Sato16} unless some mechanism stops the inward migration of the icy dust particles \citep[e.g.,][]{Morbidelli16}. 

To summarize, constraining the thermal structure of the inner regions of protoplanetary disks from astronomical observations is crucial for fully understanding how rocky planets form.  
In this article, we propose a methodology for testing disk temperature models  by multi-wavelength, high-resolution imaging with the ngVLA and ALMA.

\section{Theoretical Background}

\begin{figure*}[t]
\begin{center}
\includegraphics[width=8.5cm, bb=0 0 327 168]{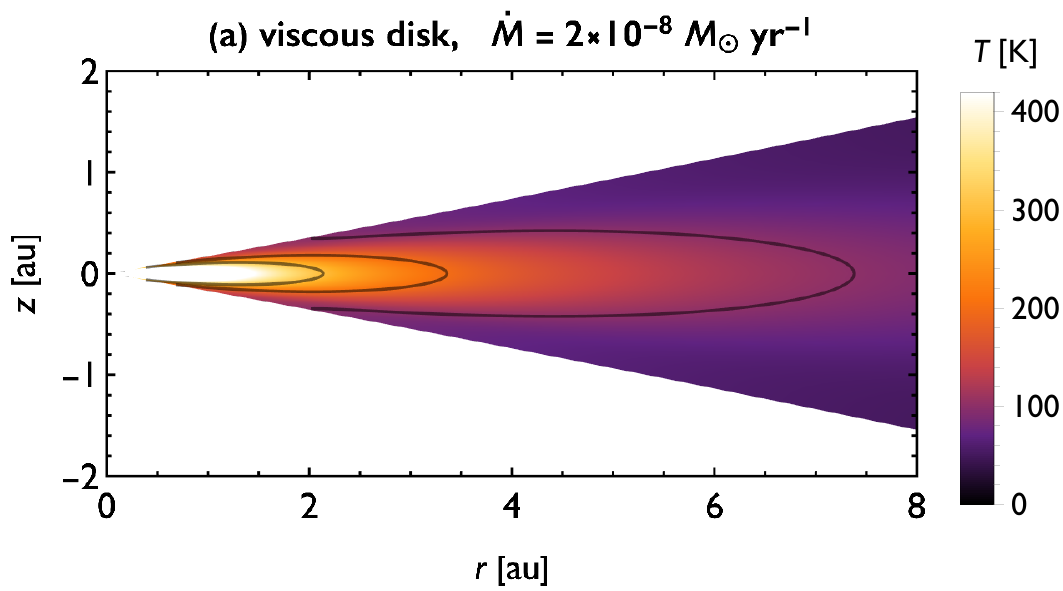}
\includegraphics[width=8.5cm, bb=0 0 327 168]{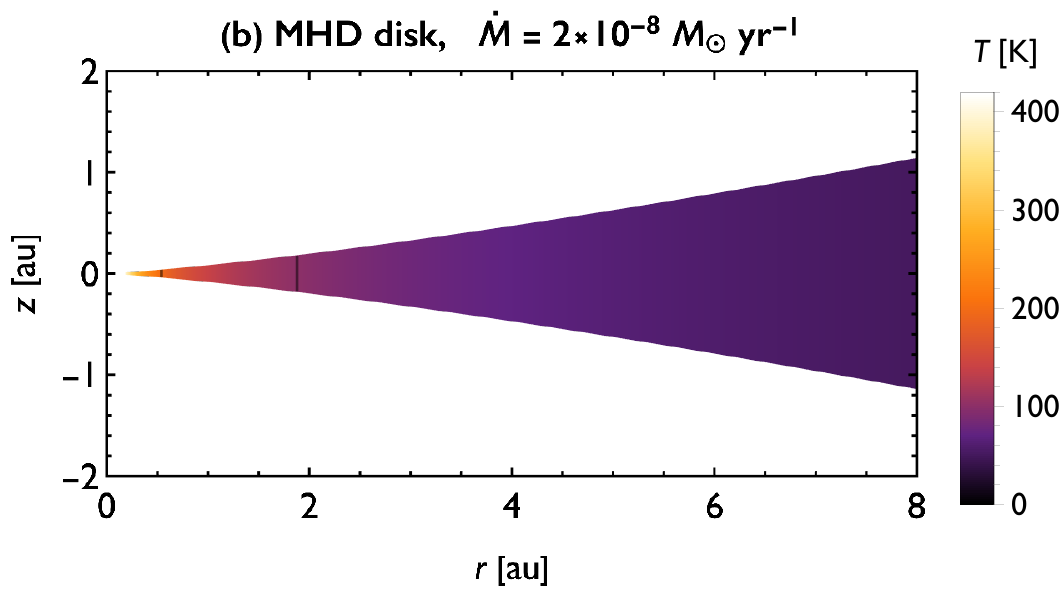}
\end{center}
\caption{Temperature as a function of the cylindrical radius $r$ and distance from the midplane $z$ for the viscous and MHD disks (panels (a) and (b), respectively).
The contours mark $T = 100$, 200, and 300 K. 
The maps are truncated at three scale heights below and above the midplane. 
}
\label{fig:T2D}
\end{figure*}

In principle, the efficiency of midplane heating can be studied by measuring disk temperatures at different vertical depths.
In classical accretion disk models with vertically uniform kinematic viscosity, accretion heating mainly takes place at the midplane, leading to the midplane temperature higher than the surface temperature in inner disk regions \citep[e.g.,][]{Davis05}.   
In contrast, in MHD accretion disk models, Joule heating takes place well above the midplane, leading to vertically uniform temperature structure with the midplane no hotter than the surface  \citep{Hirose11,Mori19}. 
This difference critically affects when the snow line crosses the 1 au orbit in the disks \citep{Mori21}. 

Multi-wavelength observations of the disks' inner regions using the ngVLA and ALMA will enable us to perform thermal tomography---imaging of the temperature structure in both the radial and vertical directions---of the rocky planet-forming regions. The ngVLA will provide access to the midplane temperature structure of the inner few au of the disks at millimeter to centimeter wavelengths, at which the inner regions are likely to be optically marginally thick to optically thin. ALMA is already capable of mapping the dust continuum emission from well above the midplane at comparable resolutions at optically thick, submillimeter wavelengths.

\section{Proof-of-concept Simulations}

\subsection{Models}

We examine the feasibility of the disk thermal tomography described above using two disk temperature models. 
One is the classical viscous disk model \citep{Lynden-Bell74} with the viscous alpha parameter $\alpha$  taken to be $10^{-3}$ throughout the disk. The other is the MHD accretion disk model recently proposed by \citet{Mori21}, which assumes accretion heating to occur on thin layers near the disk surfaces.  For simplicity, we take the depth of the heating layers to be $0.4~\rm g~cm^{-2}$ in column mass density measured from infinity \citep{Mori19,Mori21}.
The radial and vertical temperature profiles for the two accretion disk models are generated using the plane-parallel radiative transfer model by \citet{Hubeny90}. In addition to accretion heating, we also include heating by stellar radiation \citep{Chiang97}.  
For both models, we assume a uniform accretion rate of $\dot{M} = 2\times 10^{-7}~M_\odot~\rm yr^{-1}$, a stellar mass of $1M_\odot$, a stellar luminosity of $1L_\odot$, and a uniform infrared opacity of $5.0~\rm cm^2~g^{-1}$.  Figure~\ref{fig:T2D} shows the two-dimensional temperature maps for the two disk models.

The two-dimensional temperature profiles are used to generate maps of dust continuum emission from a face-on disk at submillimeter to centimeter wavelengths. 
The disk opacity in this wavelength range is assumed to follow a power law $\kappa(\lambda) = 6\times 10^{-3}(\lambda/3~\rm mm)^{-1.7}~\rm cm^{2}~g^{-1}$, where $\lambda$ is the wavelength. Again, the opacity is taken to be uniform throughout the disks, which means that we neglect possible spatial variations of the dust-to-gas mass ratio and grain size distribution due to dust evolution. 

To mimic observations, Gaussian convolution is applied to smooth the theoretical continuum intensity profiles at angular resolutions of ALMA and the ngVLA.  
Specifically, we adopt angular resolutions of 12, 5, and 16 mas for the ALMA 0.87 mm, ngVLA 3 mm, and ngVLA 1 cm bands, respectively, estimated by \cite{Ricci18} assuming 8 and 20 hr observations with ALMA and the ngVLA, respectively. The modeled disks are assumed to be located a distance of $d = 140$ pc away from the Earth, resulting in spatial resolutions of 1.7, 0.7, and 2.2 au for the 0.87 mm, 3 mm, and 1 cm band images, respectively.
Based on the sensitivity estimates by \citet{Ricci18}, the rms noise in the brightness temperature maps for the three bands are estimated to be $\approx 0.6$, 2, and 0.8 K.
The sensitivities are high enough for a precise measurement of thermal emission from around the snow line, where $T \approx 160~\rm K$. 

\subsection{Results}

\begin{figure*}
\begin{center}
\includegraphics[width=17cm, bb=0 0 683 446]{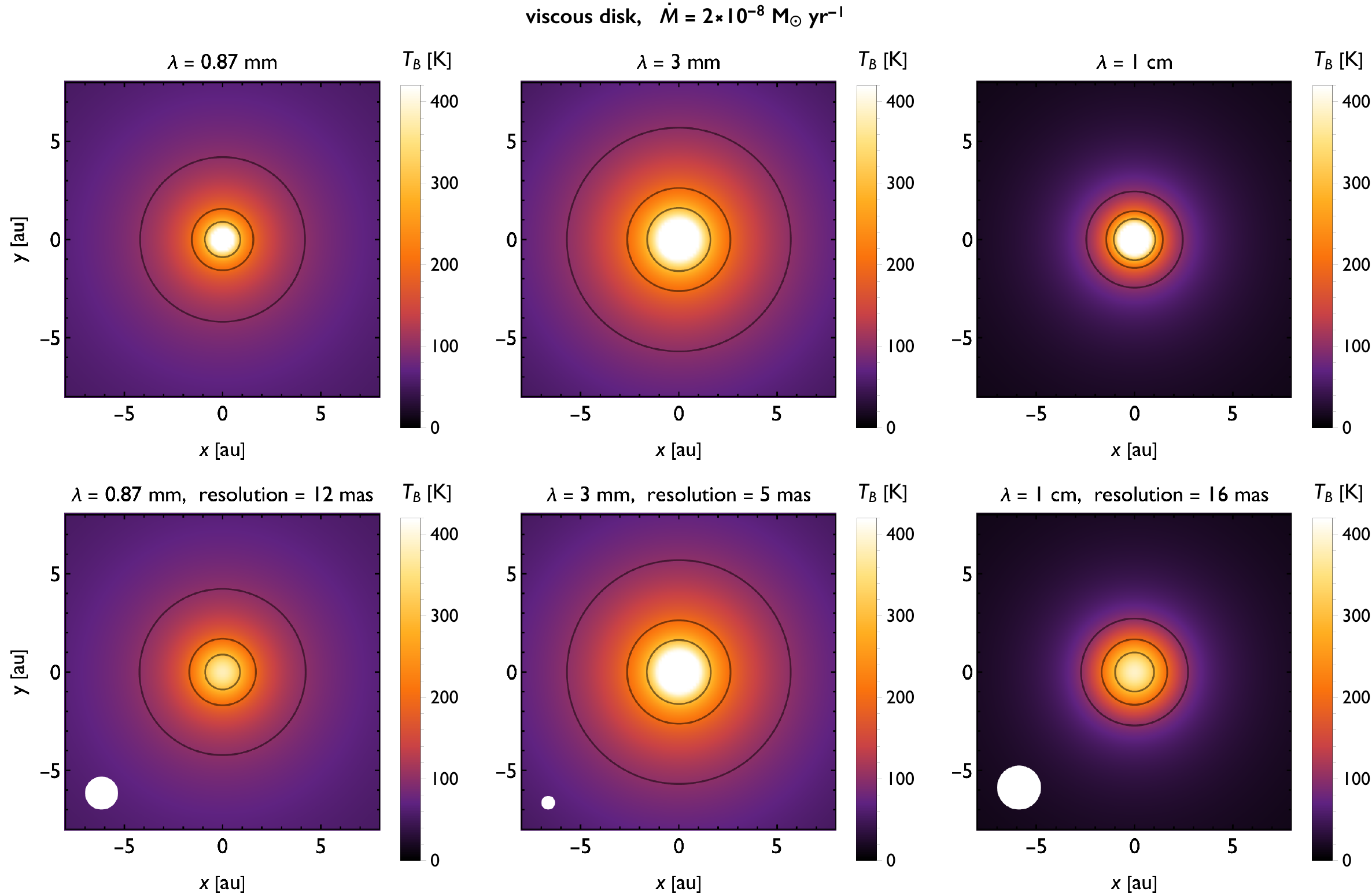}
\end{center}
\caption{Upper panels: maps of the brightness temperatures $T_{\rm B}$ at $\lambda =  0.87$ mm, 3 mm, and 1 cm (left, center, and right panels, respectively) from a face-on viscous accretion disk. The contours mark $T_{\rm B} = 100$, 200, and 300 K.
Lower panels: synthetic brightness temperature maps after Gaussian smoothing,
assuming $d = 140$ pc and spatial resolutions of 12, 5, and 16 mas for the ALMA 0.87 mm,  ngVLA 3 mm, and ngVLA 1 cm bands, respectively \citep{Ricci18}. }
\label{fig:TB_viscous}
\end{figure*}
Figure~\ref{fig:TB_viscous} presents the maps of the brightness temperatures $T_{\rm B}$ at three wavelengths from the viscous disk model. The upper and lower panels are the maps before and after Gaussian smoothing, respectively, 
Comparison between the upper and lower panels shows that the angular resolutions of the ALMA submillimeter band and ngVLA millimeter bands are high enough to resolve the radial distribution of $T_{\rm B}$ in the inner few au.   
It is interesting to note that the brightness temperature of the inner disk region is maximized  
at $\lambda = $ 3 mm and decreases both longward and shortward of 3 mm.  
This reflects the vertical thermal structure of the uniformly viscous disk, 
in which the temperature decreases as we go away from the midplane. 
At $\lambda = $ 0.87 mm, the inner region is optically thick and the thermal emission comes from 
$\approx 2$ scale heights above the midplane. At $\lambda = 3~\rm mm$, the inner region is marginally optically thick and the thermal emission is dominated by the midplane. 
Longward of 3 mm, the inner region is optically thinner at longer wavelengths, and the brightness temperature of the region decreases with increasing $\lambda$. 
This demonstrates the importance of multi-wavelength observations using ALMA and the ngVLA in studying internal heating in the inner regions of protoplanetary disks. 

\begin{figure*}
\centering
\includegraphics[width=17cm, bb=0 0 683 446]{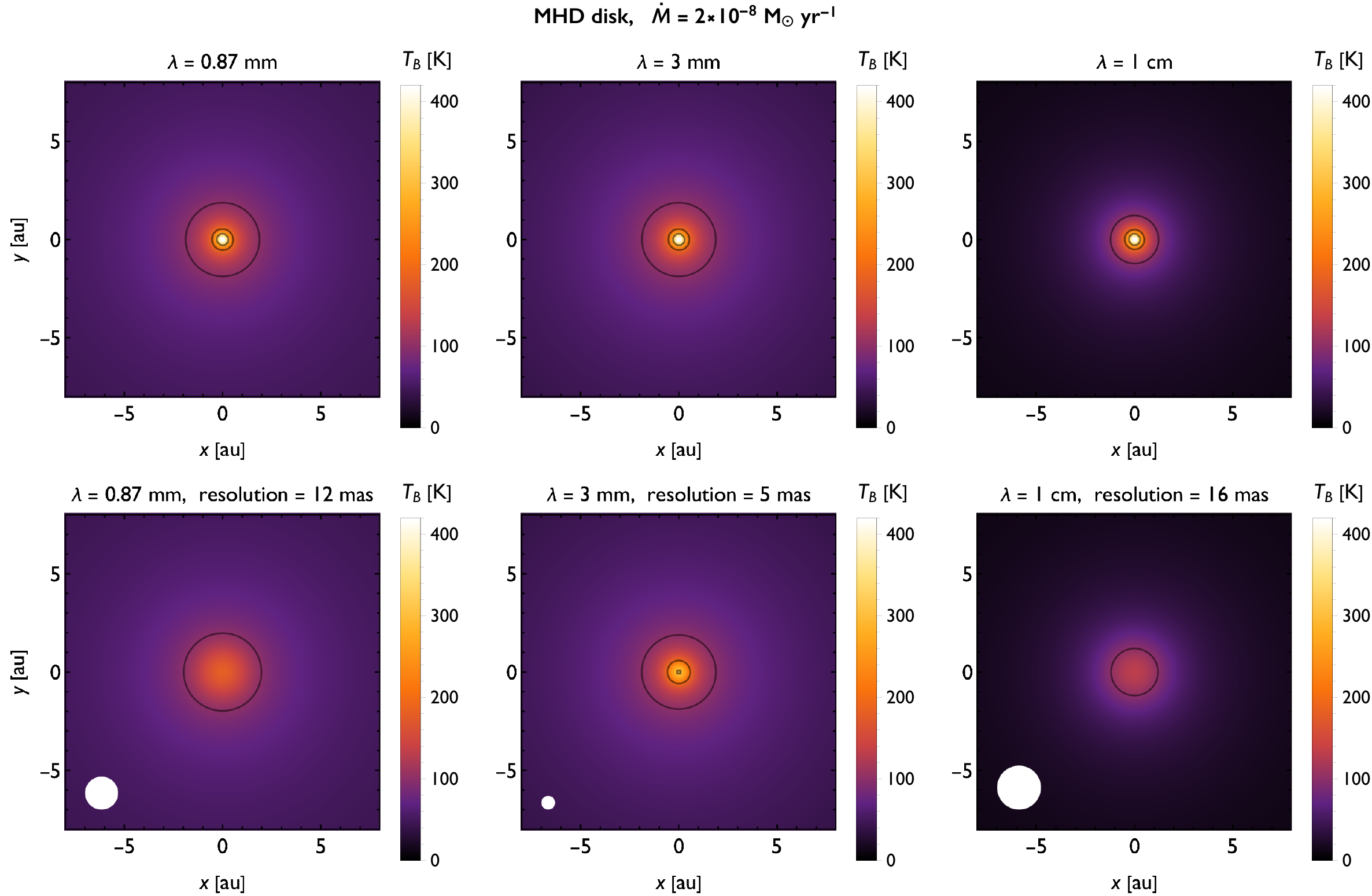}
\caption{Same as Figure~\ref{fig:TB_viscous} but for the MHD disk model. }
\label{fig:TB_MHD}
\end{figure*}
Figure~\ref{fig:TB_MHD} is the same as Figure~\ref{fig:TB_viscous}, but for the MHD accretion disk model. The brightness temperatures of the MHD disk are overall lower than those of the viscous disk. Still, the synthetic map for the ngVLA $3~\rm mm$ band accurately captures the brightness temperature profile of the MHD disk up to $T_{\rm B} \approx 200~\rm K$ and thus allows us to study the thermal structure around the water snow line ($T \approx 160~\rm K$). Because of its vertically uniform temperature structure, the MHD disk model produces nearly identical brightness temperature maps (except for the unresolved very inner region) at optically thick wavelengths of $\lambda \lesssim 3~\rm mm$. Therefore, multi-band observations at ALMA submillimeter bands and the ngVLA 3 mm band would allow us to test whether midplane heating is inefficient around the midplane in a disk.
 
\section{Conclusion}
We have demonstrated that thermal tomography with the ngVLA and ALMA will enable us to discriminate between different models for disk thermal structure. 
Such observations will provide us with a basis for fully understanding when and where terrestrial planets form. 
However, the models we have used here neglect a number of important dust evolution processes, including coagulation, vertical sedimentation, and radial drift. In particular, condensation, sublimation, and sintering of icy particles can significantly affect the size and spatial distribution of solids around the snow line \citep[e.g.,][]{Birnstiel10,Ros13,Okuzumi16,Schoonenberg17}. All these dust evolution processes may alter the temperature structure and emission profiles in the inner regions of protoplanetary disks \citep[e.g.,][]{Banzatti15,Pinilla17} and therefore should be included in future modeling.

\section{Acknowledgments}

We thank Takahiro Ueda for discussions. This work was supported by JSPS KAKENHI Grant Numbers JP18H05438, JP19K03926, JP19K03941, and JP20H00182. 
 

\end{document}